\documentclass[fleqn,10pt]{wlscirep}
\usepackage[utf8]{inputenc}
\usepackage{multirow}
\usepackage[T1]{fontenc}
\title{Semantic Segmentation Based Quality Control of Histopathology Whole Slide Images}

\author[1,*]{Abhijeet Patil}
\author[2,3]{Garima Jain}
\author[1]{Harsh Diwakar}
\author[1]{Jay Sawant}
\author[4]{Swapnil Rane}
\author[4]{Tripti Bameta}
\author[3]{Sanghamitra Pati}
\author[1]{Amit Sethi}
\affil[1]{Indian Institute of Technology - Bombay, Department of Electrical Engineering, Mumbai, 400076, India}
\affil[2]{Indian Institute of Technology - Bombay, Koita Centre for Digital Health, Mumbai, 400076, India}
\affil[3]{Indian Council of Medical Research, National Institute for Research in Digital Health and Data Science, Delhi, 110029, India}
\affil[4]{Tata Memorial Centre -- ACTREC, HBNI, Kharghar, Navi Mumbai, 400011, India}

\affil[*]{abhijeetptl5@gmail.com}

\begin{abstract}
We developed a semantic segmentation models for quality control (QC) of histopathology whole slide images (WSIs) that segments various regions, such as blurs of different levels, tissue regions, tissue folds, and pen marks. Given the necessity and increasing availability of GPUs for processing WSIs, the proposed models comprises multiple lightweight deep learning models to strike a balance between accuracy and speed.  These models were evaluated on all TCGA WSIs, which is the largest publicly available WSI dataset containing more than 11,000 histopathological images from 28 organs. It was compared to a previous work, which was not based on deep learning, and it showed consistent improvement in segmentation results across organs. To minimize annotation effort for tissue and blur segmentation, annotated images were automatically prepared by mosaicking patches (sub-images) from various WSIs whose labels were identified using a patch classification model. Due to the generality of our trained QC models and its extensive testing the potential impact of this work is broad. It can be used for automated pre-processing any WSI cohort to enhance the accuracy and reliability of large-scale histopathology image analysis for both research and clinical use. We have made the trained models, training scripts, training data, and inference results publicly available at~\href{https://github.com/abhijeetptl5/wsisegqc}{Github URL}, which should enable the research community to use the models right out of the box or further customize it to new datasets and applications in the future.
\end{abstract}
\begin{document}

\flushbottom
\maketitle

\thispagestyle{empty}

\section*{Introduction}

\label{sec:introduction}
The practice of pathology is undergoing a digitization process worldwide based on the adoption of whole slide scanners along with digital whole slide image (WSI) storage, management, and archiving solutions. Digital pathology not only supports telepathology and virtual tumor boards, but it also promises to enhance diagnostic accuracy, precision, and throughput with the application of computational pipelines based deep learning models and image processing algorithms. Such pipelines have shown promise in various problems, including detecting and classifying tumors in WSIs~\cite{clam}, grading tumors~\cite{gleason}, and predicting patient outcomes~\cite{digipathconcordance3, her2_garima}.

These pipelines, however, often falter when faced with the variability inherent in routine clinical practice. Factors such as differences in slide preparation and staining protocols, variability in technician training and experience, and the use of diverse scanner models and imaging settings can introduce subtle shifts in color, focus, and overall image quality. In research settings, dedicated funding and rigorous quality assurance ensure consistently high‐quality scans, but those safeguards are seldom available in high‐volume diagnostic laboratories. As a result, computational models trained under idealized conditions may produce unreliable results on real‐world whole slide archives. To achieve the consistency and reliability needed for clinical adoption, it is therefore critical to implement comprehensive quality‐control measures that monitor every stage of the digital pathology workflow, from slide preparation through image acquisition and archiving~\cite{stresstest, jbhieffect}.

\begin{figure}
	\centering
	\includegraphics[width=0.9\textwidth]{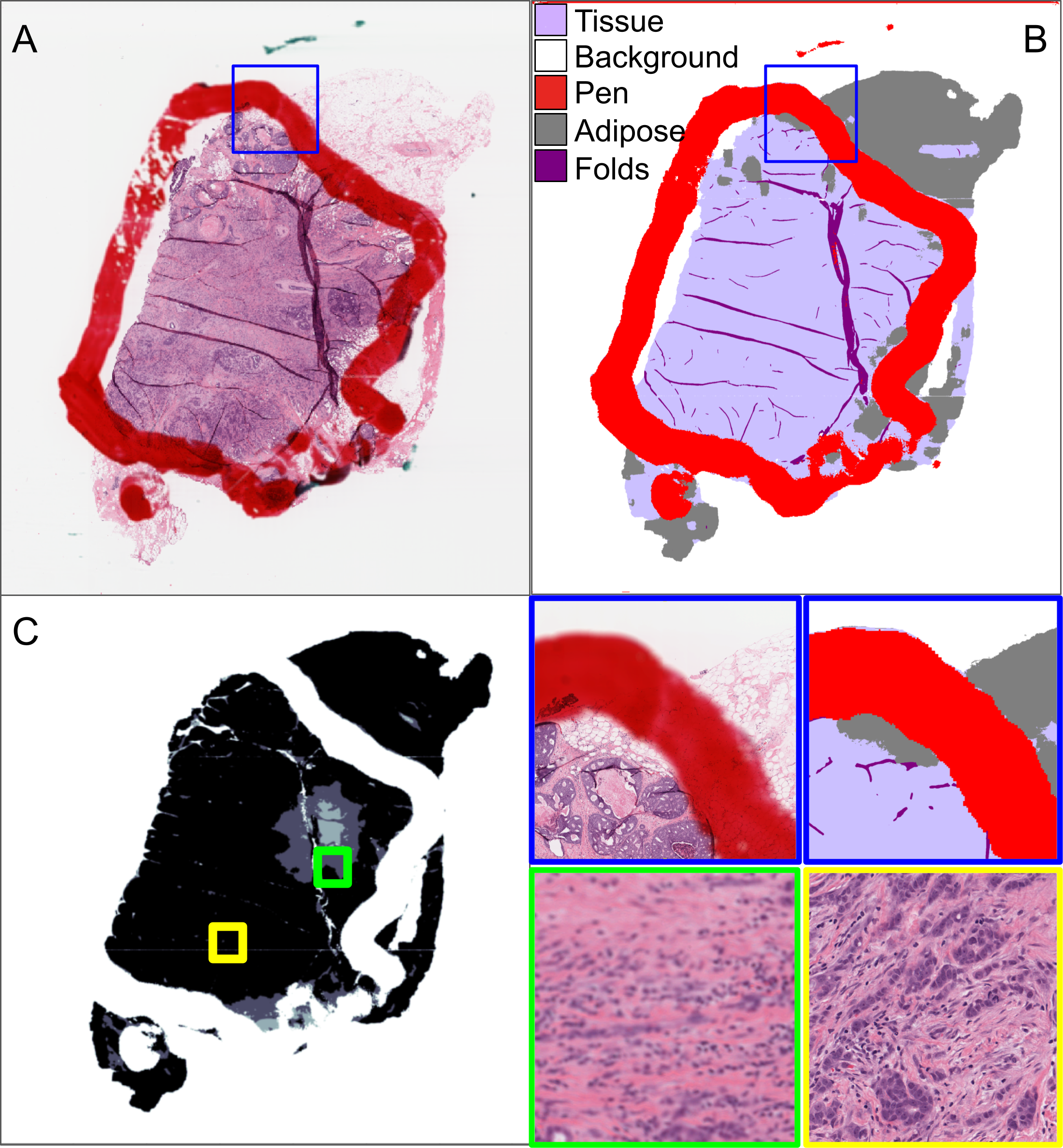}
	\caption{\textbf{Preview of results:} (A) WSI; (B) predictions of models that detects tissue (lilac), adipose (gray), pen mark (red), and tissue folds (purple), respectively; (C) predictions of blur-level detection model (brighter mask regions indicate more blur); and (D) a few zoomed in patches to show image details (boundary color coded to show their location in WSI).} 
	\label{fig:result_preview}
\end{figure}

In this study, we introduce hand-annotated datasets and a novel technique for training segmentation models for histopathology. The datasets include annotations for pen marker segmentation and tissue fold segmentation, and we have developed four segmentation models for blur level segmentation, tissue segmentation, tissue fold segmentation, and pen marker segmentation. Our study aims to address the challenges associated with histopathology image analysis, particularly in identifying different tissue components and artifacts. The hand-annotated datasets and segmentation models we introduce can assist in improving the accuracy and efficiency of histopathology image analysis. Sample predictions from our models is shown in Figure \ref{fig:result_preview}.

To develop our models, we utilized a novel technique for training segmentation models, which takes into account the unique characteristics of histopathology images. The four segmentation models we developed were designed to address different aspects of histopathology image analysis, such as identifying tissue components, blurry regions, tissue folds, and pen markers. We conducted a thorough analysis of the largest publicly available dataset to validate the accuracy and effectiveness of our models. Main contributions of our work are
\begin{itemize}
    \item We present a novel method to train a segmentation model for histopathology images which utilizes domain knowledge generated from HistoROI~\cite{abhijeetjpi} for better sampling of data.
    \item We make hand-annotated datasets for pen marker segmentation and tissue fold segmentation publicly available. We release our training and inference codes along with weights of trained models publicly available. These datasets and models can help the research community to develop better and efficient quality control tools.
    \item We also release the predictions of our models on more than 11,000 WSIs in TCGA data portal~\cite{tcga}. These predictions can be used to develop weakly supervised learning algorithms for quality control of histopathology images. Also, researchers can utilize these predictions to prepare and choose their datasets keeping quality of images in check.
\end{itemize}

In this section, we give an overview of studies related to quality control in histopathology images. These studies can be categorized into two broad categories: traditional image processing-based methods and deep learning-based methods. Most of the studies focus on addressing specific types of tissue defects, while some attempt to solve multiple defects with a single model or separate models.  HistoQC~\cite{histoqc} is by far the most widely used histopathology QC (Quality Control) tool. It performs multiple tasks using conventional image processing, machine learning techniques and deep learning to identify regions such as useful tissue region, adipose-like region, background, tissue folds, out-of-focus region, etc. It also provides an easy to use HTML (HyperText Markup Language) based UI (User Interface) to adjust parameters depending on staining protocols used for slide preparation. It has been shown that HistoQC can identify foreground regions correctly for more than 95\% WSIs\cite{histoqc}. 

A wide variety of work, inspired from natural image settings has been applied to pathology images to identify out-of-focus regions~\cite{towardsmlqc}. This problem has seen a lot of application in the research community as well as in industrial settings. Identification of lens parameters is of paramount importance while capturing images through digital scanners, which can be guided by automatic focus assessment algorithms. It has been shown that by comparing various quality metrics that are used for natural images, such as SSIM (Structural Similarity Index Measure) ~\cite{imagequalityssim} and IL-NIQE (Integrated Local - Naturalness Image Quality Evaluation) ~\cite{blindimagequality}, sharply captured histopathology image patches can be distinguished from patches that are out-of-focus~\cite{automaticqualitypath}. A few studies have experimented with several full reference and no-reference image quality metrics and validated these metrics with pathologists~\cite{qualitymetricspath}. Image gradient based features have also been used for identification of out-of-focus regions in WSIs~\cite{automatedblur}. Few studies have trained machine learning based algorithms such as linear regression~\cite{referencelessquality}, random forest~\cite{towardsmlqc}, etc. on top of image features to determine focus quality. Synthetically blurred images were also used in few studies to build classifiers to detect out-of-focus patches~\cite{automateddatagen}. Computationally-efficient kernels based on the human visual system, e.g. HVS (Human Visual System) - MaxPol, have also been utilized to detection out-of-focus regions~\cite{focusqualityassessment}. Convolutional Neural Network (CNN) based deep neural networks have also been employed for out-of-focus region detection tasks~\cite{deepfocus, focusordinal}. Light weight neural networks have also shown capability to distinguish between sharp and out-of-focus images with high accuracy~\cite{focuslitenn}.

There are relatively fewer studies concentrating on other types of defects compared to out-of-focus artifacts. A major challenge in addressing these defects is unavailability of annotated datasets and difficulty in creating them. These defects are also hard to generate synthetically. A few studies dealing with tissue fold segmentation utilize simple features, such as color and connectivity properties of tissue structures color saturation and luminance, to detect tissue folds~\cite{tissuefoldpred}. A few studies have also explored feature selection for distinguishing patches with tissue fold to be used by machine learning algorithms for classification~\cite{tissuefold2}. Deep learning based classifiers have also been used to classify patches with tissue fold~\cite{tissuefold3, tissuefold4}.

For tissue segmentation tasks, few studies have proposed CNN based segmentation models. Most of the studies used metrics like R/B ratio, pixel intensity, etc~\cite{clam, abhijeetjpi} for tissue segmentation. Generative Adversarial Networks (GANs) are being used to identify out-of-distribution patches as a proxy to segregate pen marker affected regions in WSIs~\cite{inkremove}. Systematic analysis of stains also uncovered potential to identify pen marker affected regions~\cite{slidenet}.

Many studies try to identify problematic artifacts affected regions without segregating each type of artifact as a separate category. Approaches like supervised learning~\cite{pathprofiler}, weakly supervised learning~\cite{artifactidentificationinwsi}, few-shot learning~\cite{artifactidentificationfewshot}, learning with noisy labels~\cite{snowsupervision}, active learning~\cite{abhijeetjpi}, explainable AI (Artificial Intelligence) ~\cite{nikhilrobust}, etc. have been explored in this space. A few studies have also touched upon the importance of quality control, showcasing increase in WSI classification performance for various datasets. Reviews of QC techniques  for histopathology images can also be found~\cite{qcreview1, qcreview2}.

\section*{Results}

\subsection*{Comparison with HistoQC -- WSI level}
\begin{figure}
	\centering
	\includegraphics[width=0.8\textwidth]{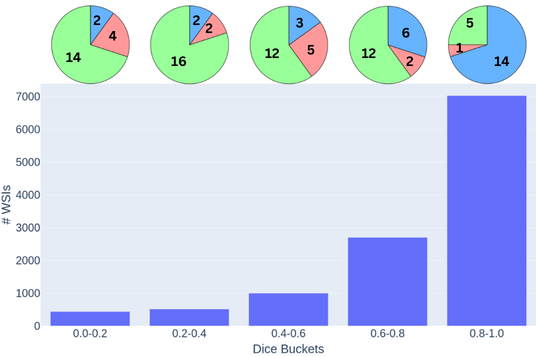}
	\caption{Agreement between HistoQC and our models for 11,650 WSIs in TCGA dataset. Bar charts shows distribution of WSIs according to Dice score agreement between our models and HistoQC. Pie charts show subjective comparison on 20 WSIs sampled from each bucket of agreement by pathologist between masks predicted by our models and HistoQC (Green - Our better, Red - HistoQC better, Blue - inconclusive). Our models perform better than HistoQC.}
	\label{quantitative_results}
\end{figure}

To evaluate the performance of our models, we compared it with HistoQC, a widely used quality control tool. We ran all 11,666 WSIs available on the TCGA data portal through HistoQC's default parameters to generate a foreground tissue mask, and performed a similar procedure to predict foreground masks through our proposed models. We then computed the Dice score between the masks to measure the agreement between the two predictions. A high Dice score indicates a high agreement between HistoQC and our models, indicating that the masks generated by both strategies are either equally  good or equally bad. However, in cases of lower agreement (low Dice score), there may be instances where HistoQC performs better, our models perform better, or it may not be possible to determine which mask is better. Figure \ref{quantitative_results} summarizes findings from this experiment.

We sorted the Dice scores of the foreground tissue masks generated by HistoQC and our models and divided them into buckets according to the Dice scores. We sampled 20 WSIs from each bucket, resulting in a total of 100 WSIs, which were then manually analyzed for quality by pathologists. Proportion of WSIs in  each bucket is observed to be 60\%, 20\%, 10\%, 5\%  and 5\% for  the bins of dice 0.8 to 1, 0.6 to 0.8, 0.4 to 0.6, 0.2 to 0.4 and 0  to 0.2  respectively. The pathologists were presented with two masks in QuPath\cite{qupath} for each WSI without being told how these masks were generated. For each pair of masks, the pathologists were asked to select the better one. The order of the two masks for each WSI was randomized.

From the bucket of the most agreement, the outcome of comparison to most of the WSIs was not conclusive, which was  expected because of a higher Dice score. Out of 20 WSI mask pairs in this bucket,  comparison for 14 was inconclusive, for 5 WSIs our models  performed better than HistoQC and  for one WSI, HistoQC performed better.  Results were more conclusive on the other end, where our models performed better than HistoQC for 14 and 16 WSIs for the dice bucket of 0.2 to 0.4 and 0 to 0.2  respectively. A few samples were inconclusive in these buckets because both the models performed equally bad for few WSIs. Though our predictions are better than HistoQC, comparison  based on WSIs predictions becomes a hectic task and can add too much subjectivity and bias while comparing different masks. Therefore we have carried out patch level mask comparison to address minute differences in HistoQC and our models.

\subsection*{Comparison with HistoQC -- Patch level}
\begin{figure}
	\centering
	\includegraphics[width=0.8\textwidth]{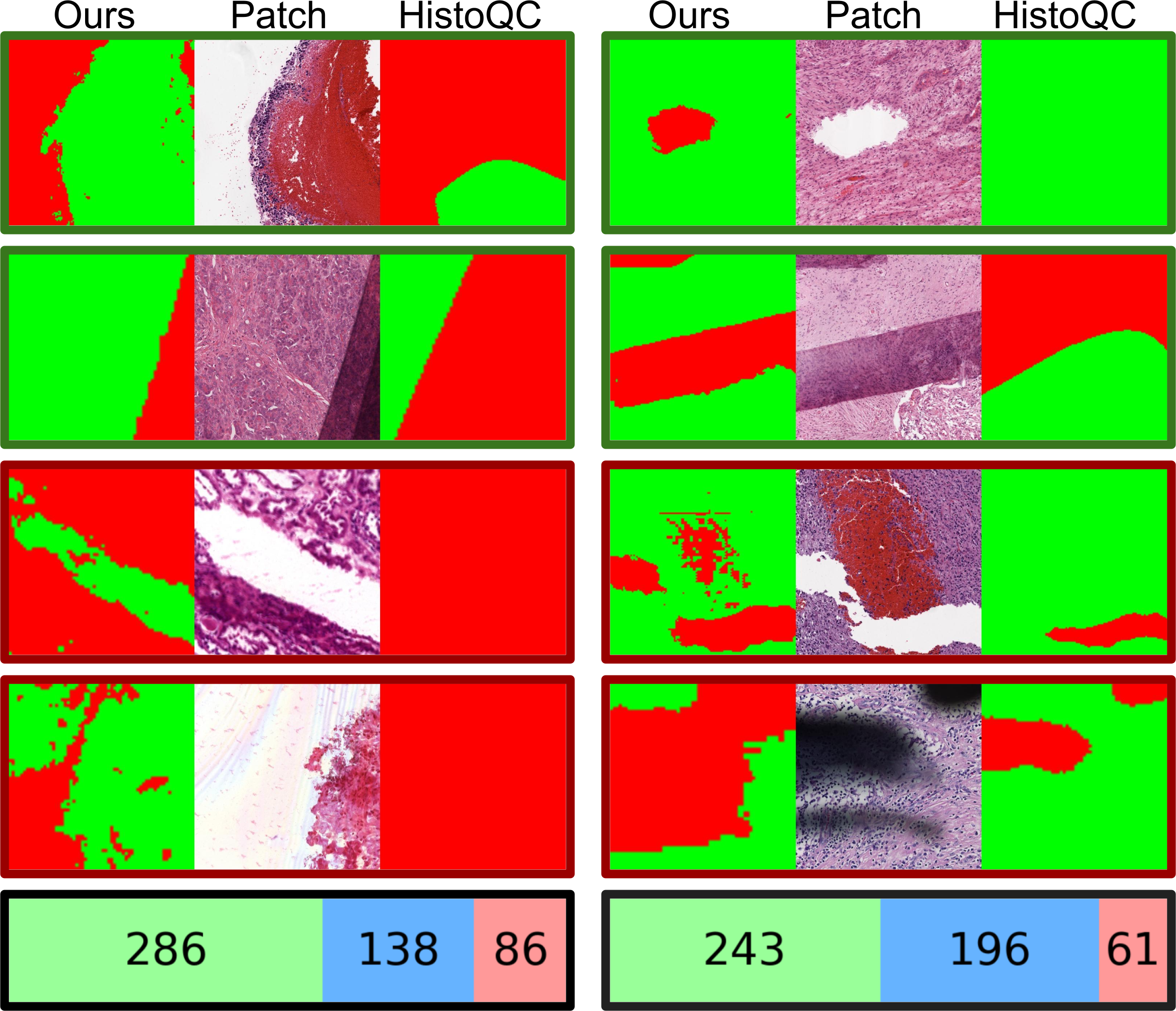}
    	\caption{Comparative Analysis of Model Performances Against HistoQC. Left half shows comparison for SET-1 and right half for SET-2. For all masks, green color indicates foreground and red indicates background. For patches marked with green border, our mask is subjectively better than HistoQC and for red border, HistoQC mask is better. Bottom bar chart shows statistics for overall comparison on SET-1 and SET-2 (Green-Ours better, Red-HistoQC better, Blue-Non-conclusive). For both sets, our model performs better than HistoQC.}
	\label{quantitative_results_patch}
\end{figure}

In this experiment, we have systematically  sampled patches from WSIs along with foreground masks predicted by our models and by HistoQC. We have selected 100 WSIs from the TCGA dataset for extracting patches in  this experiment. We first identify the regions of disagreements between both the masks, and then sample 5 patches from a region where HistoQC predicted a foreground but our models predicted background and vice-versa. With this strategy, we sample 10 patches from each of 100 WSIs, generating a dataset of 1,000  patches along with their predictions by HistoQC and our models. We call a set of 500 patches for which the center pixel of prediction is foreground for HistoQC but background for our models, SET 1 and for another scenario, SET 2. Each of these sets contains 500 patches. Results from this experiment is shown in Figure \ref{quantitative_results_patch}.

Out of 500 patches from SET 1, comparison for 138 patches is non conclusive. From the remainder of 362 patches, our models performed better on 286 patches and HistoQC performed better on 86 patches. In the majority of cases, HistoQC can not identify tissue folds as background but our models can predict tissue folds correctly.  On the other hand, our model predicts background for debris, etc, which HistoQC correctly predicts as foreground. Analysis of SET 2 uncovers false positives predicted by our models for background class. Overall, performance of our models is better than that of HistoQC in this set as well. Out of 500 patches, 196 were not conclusive, whereas 243 patches from our models were observed to be better than HistoQC. Out of 61 patches where HistoQC performed better, most of these patches correspond to small bubbles or wipes on glass slides.

\section*{Discussion}

In this study, we introduced a quality control models for histopathology images using four different segmentation models, enhanced by HistoROI predictions and the image collage method to streamline tissue and blur segmentation. Our approach demonstrates superior performance on the largest publicly available dataset, and we have made all models, scripts, and results freely available to foster further advancements in this field. Our models offer a promising method for improving histopathology image analysis, with significant potential for future enhancements that could lead to more accurate and efficient techniques.

\section*{Methods}

The accurate identification and classification of artifacts is essential for effective analysis of WSIs. In this study, we developed four distinct segmentation models—each tailored to a specific task: blur‐level estimation, tissue‐fold detection, pen‐marker identification, and tissue‐foreground segmentation. Although the tissue segmentation model does not target an “artifact” per se, it is indispensable for quality control: by precisely delineating diagnostically relevant tissue regions from non‐tissue background, it ensures that our artifact detectors operate solely within meaningful areas, thereby reducing false positives and streamlining downstream workflows. Moreover, we explicitly treated adipose as a separate tissue class, since its homogeneous, low‐texture appearance can easily be misinterpreted as blurred artifact by an inadequately trained blur model. We utilized annotated datasets for pen‐marker and fold segmentation, and an image‐collage approach for both blur and tissue segmentation, leveraging HistoROI predictions to generate varied, realistic training examples~\cite{abhijeetjpi}. In the following sections, we detail our model architectures, the training and validation datasets, and the optimization strategies—including data augmentation and fine‐tuning—employed to maximise performance.

\subsection*{Image collage method}
Training segmentation models for WSI level can be a challenging task as it requires obtaining a large amount of image annotations, which take substantial time and effort. Even if annotated data is available, the data distribution is often skewed, which can lead to a biased model. Training deep neural networks requires each batch of data to contain representative data, but this is not always possible with WSIs since ROIs generally span thousands of pixels.

To address this challenge, we have used a novel image collage method for training segmentation networks. This method involves copying patches with known labels (corresponding to the majority of their pixels) from WSIs and pasting them together with patches from other WSIs after downscaling fixed size to form larger images composed of diverse patches. The annotation masks of these larger images are like pixelated grids, where pixels in each cell of the grid have the same annotation label. This approach not only makes it easy to obtain annotation masks for training segmentation models, but also ensures that each training batch includes representative samples to address the class imbalance in WSIs.

We have used the HistoROI patch classification model to assist batch formation in the image collage method. HistoROI model predicts the class of a patch of a WSI, which can be epithelial, stroma, lymphocytes, adipose, miscellaneous, or artifact. By using this model, we can select patches or regions from specific classes, which ensures that the balance between different types of tissue regions is maintained during the training of the segmentation model. The HistoROI model uses a deep neural network trained on patch-level data, which makes it highly accurate in predicting the class of a patch, but it is not trained for segmentation. By combining the image collage method and HistoROI model predictions, we were able to overcome the challenge of lack of annotated data and skewed data distribution. This allowed us to train highly accurate segmentation models for WSI level tasks. 

\subsection*{Tissue segmentation model}
\begin{figure}
	\centering
	\includegraphics[width=0.9\textwidth]{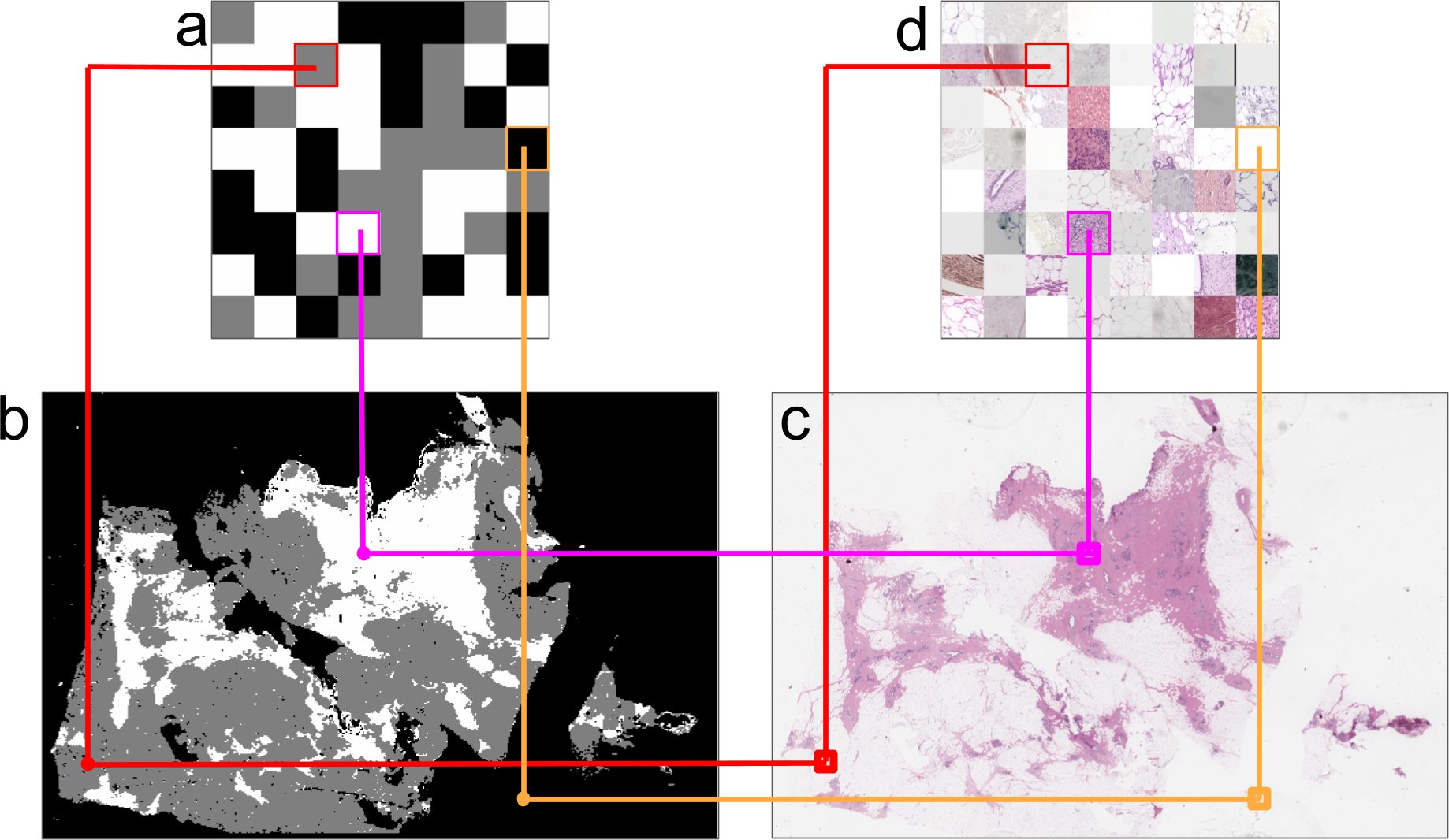}
	\caption{\textbf{Image collage method} to generate uniform data distribution across all classes. (a) A grid of patches of size 64x64 placed on the empty array of size 512x512. (b),(c) A WSI alongwith its HistoROI predictions is sampled from a pool of WSIs selected from training dataset. (d) A patch of required class is randomly selected from WSI. Index of patch is obtained by HistoROI predictions. Patches from 2.5X magnification are used to train this model. A CNN segmentation model is trained using randomly generated masks and images created by pasting patches from WSI. Generated images can have patches from multiple WSIs, making the segmentation model more robust.}
	\label{figure_cut_paste_method_tissue}
\end{figure}

The training began with dataset preparation, in which empty 2D arrays of 512 × 512 pixels were divided into an 8 × 8 grid of 64 × 64-pixel cells. Each cell was randomly assigned one of three labels—foreground tissue, adipose, or background—to generate annotation masks. For every grid cell, patches were mined from a pre-selected pool of twenty whole-slide images per class, ensuring sufficient variation to support accurate model training. These patches were extracted at 2.5× magnification, striking a balance between high spatial resolution and computational efficiency. The overall data-generation workflow is illustrated in Figure \ref{figure_cut_paste_method_tissue}.

This segmentation strategy not only utilized the precise predictions produced by HistoROI but also introduced variability by sampling patches from diverse slide sources. It ensured that each class was evenly represented, which in turn made the model more robust and capable of generalizing to unseen data. Adipose tissue was explicitly treated as its own class—separate from the general background—to address the specific challenge of distinguishing non-diagnostic fat regions from other background elements such as coverslip artifacts or pen markings.

Initially, we applied color‐normalization techniques~\cite{abhijeetclrnorm, vahadane} to mitigate staining variability, but subsequent experiments showed that models trained with color‐jitter augmentation consistently outperformed those relying on normalization. Moreover, the normalization step introduced significant computational overhead, making it impractical for large‐scale WSI processing. Consequently, we omitted color normalization and relied solely on HistoROI-guided patch selection, strategic patch mining, and color-jitter augmentation. This combination produced a balanced and diverse training set that significantly enhanced tissue‐detection performance on whole-slide images.

\subsection*{Blur level segmentation}
Detecting blur levels in whole-slide images proved challenging because tissue texture varied markedly across regions—for instance, stromal areas appeared smoother than cellular regions. Traditional Laplacian‐based methods often failed to deliver consistent results under these conditions. To address this, we leveraged HistoROI‐guided patch mining, which enabled the model to learn the characteristic texture of each region type directly from the data. By sampling patches that HistoROI had identified as belonging to specific tissue classes, the segmentation network acquired a more nuanced understanding of regional textures, thereby overcoming the limitations of purely gradient-based blur detectors. This synthetic data strategy, underpinned by HistoROI’s predictions, was particularly valuable in scenarios where manually annotated blur examples were scarce or unavailable.

\begin{figure}
	\centering
	\includegraphics[width=0.9\textwidth]{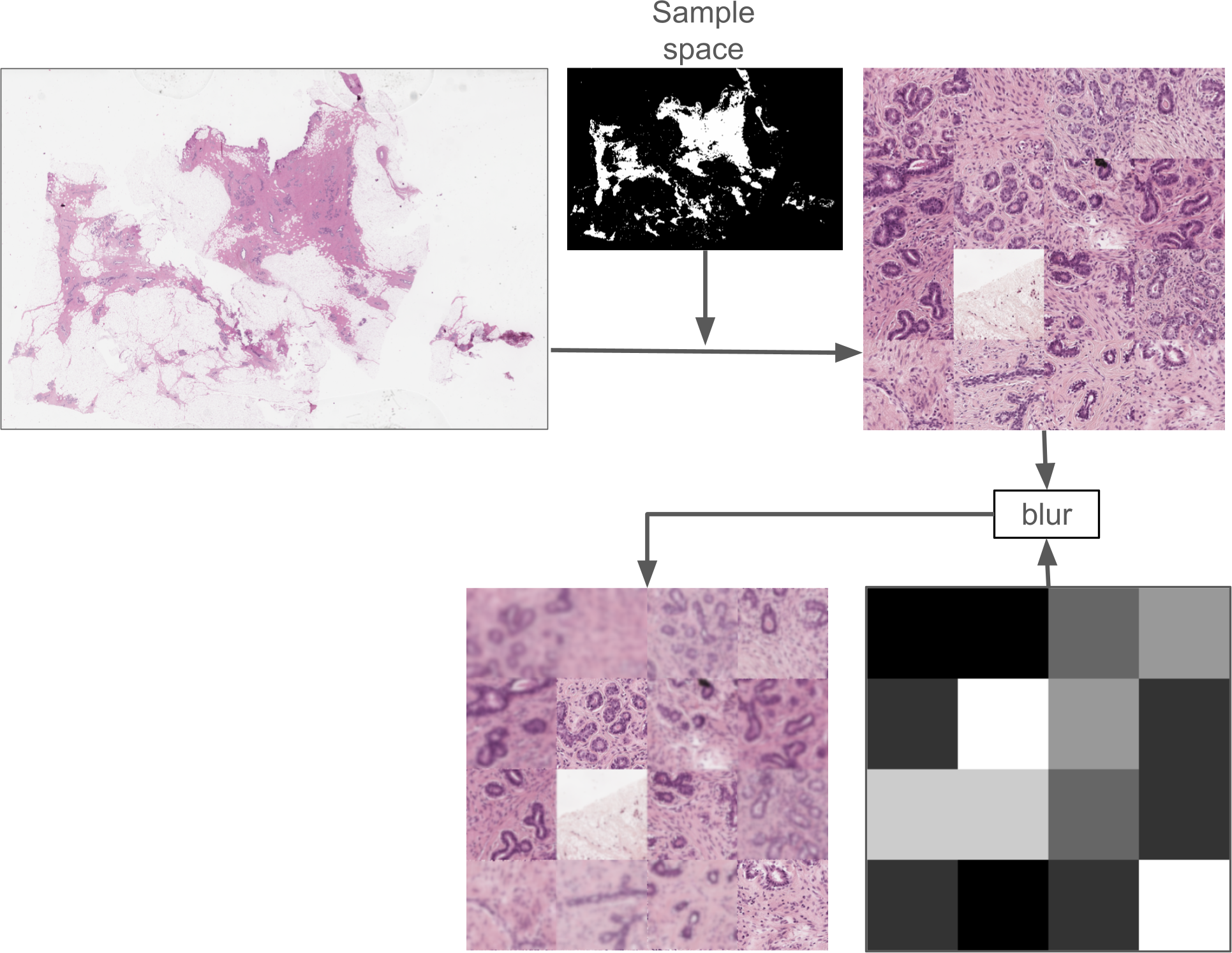}
	\caption{\textbf{Image collage method} for training blur segmentation model.}
	\label{figure_cut_paste_method_blur}
\end{figure}

\begin{figure}
	\centering
	\includegraphics[width=0.8\textwidth]{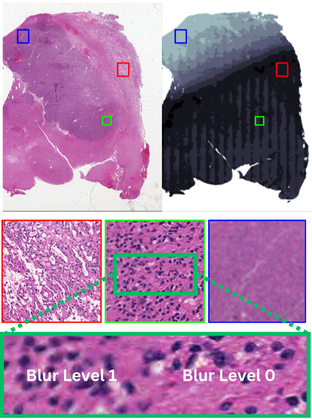}
	\caption{\textbf{Blur Segmentation model prediction:} Model predicts no blur for red patch. Blue patch shows out of focus region, correctly predicted as blur (class 7) by blur segmentation model. Green patch shows capability of our model to detect subtle blur artifacts due to potential scanner calibration issues.}
	\label{figure_blur_model_predictions}
\end{figure}

To construct the blur‐level segmentation task, we first applied the PIL BoxBlur filter to foreground patches extracted at 5× magnification, creating a continuum of blur intensities. Each 512 × 512 mask was partitioned into a 4 × 4 grid, and individual cells were randomly assigned one of eight blur classes (Class 0 for no blur through Class 7 for maximal blur). The corresponding input images were generated by blending HistoROI predictions with progressively stronger BoxBlur parameters. This procedure yielded paired inputs and ground-truth masks that defined an eight-class segmentation problem (see Figure \ref{figure_cut_paste_method_blur}), and sample model outputs are illustrated in Figure \ref{figure_blur_model_predictions}.

Finally, we experimented with applying blur operation at various magnifications. In the final model, we sampled 1024 × 1024 patches at 40×, applied BoxBlur at seven incremental levels, and then downsampled each blurred patch to 128 × 128—equivalent to 5× magnification. Although direct training at 40× would have been ideal for fine‐grained blur estimation, the computational burden proved prohibitive given the size of whole-slide images. Blurring at high resolution before downsampling preserved subtle focus artifacts while maintaining a manageable workload. This approach struck a balance between image fidelity and efficiency, enabling our model to detect both pronounced and subtle blur discrepancies arising from scanner calibration errors.

\subsection*{Tissue Folds and Pen marker segmentation}
We trained segmentation models to detect pen markers and tissue folds on whole-slide images using a consistent sampling strategy. For pen-marker detection, we randomly selected a positive pixel from each annotation mask and extracted the surrounding 512 × 512 patch as input, ensuring the model was exposed to a wide variety of marker colors and thereby enhanced its generalization. A comparable approach was applied to the tissue-fold dataset, where folds were identified and patches mined in the same manner. Both models were developed and evaluated using an 80\%–20\% training–validation split to provide a balanced assessment of their performance.

\subsection*{Model architectures}
The four segmentation models under consideration have been evaluated using two architectures: UNet~\cite{unet} and UNet++~\cite{zhou2019unetplusplus}, each with multiple backbones, including ResNet18~\cite{resnet}, ResNet34, and EfficientNet-b0~\cite{efficientnet}. These backbones are pre-trained on the ImageNet dataset, ensuring the models can effectively extract useful features from input images. We have utilized PyTorch, a popular deep learning framework, for training these models, along with the segmentation-models-pytorch library, which is available via PIP. This library offers a range of pre-defined architectures and loss functions commonly used in image segmentation tasks, simplifying the implementation process.

Model performance and inference time for each model are shown in the Table \ref{table_seg_model_compare}. Considering both execution time and accuracy, we have selected the UNet with ResNet18 backbone for tissue fold, blur, and tissue segmentation tasks. For pen marker segmentation, we have chosen the UNet++ with ResNet34 backbone. Although UNet++ with the ResNet34 backbone performs better for all tasks, its inference time is almost three times that of UNet with the ResNet18 backbone. However, we opted for UNet++ with ResNet34 for pen marker segmentation because this model uses thumbnails at a 0.625X magnification level for predictions. The overall impact of the pen marker model on the inference time for combined model inference is not significant. Inference times for each combination of tried models for all the tasks is given in Table \ref{table_seg_model_compare}.

\bibliography{sample}

\section*{Author contributions statement}

Must include all authors, identified by initials, for example:
A.A. conceived the experiment(s),  A.A. and B.A. conducted the experiment(s), C.A. and D.A. analysed the results.  All authors reviewed the manuscript. 

\section*{Additional information}

To include, in this order: \textbf{Accession codes} (where applicable); \textbf{Competing interests} (mandatory statement). 

The corresponding author is responsible for submitting a \href{http://www.nature.com/srep/policies/index.html#competing}{competing interests statement} on behalf of all authors of the paper. This statement must be included in the submitted article file.

\begin{table}
\begin{tabular}{|c|cc|cc|cc|cc|}
\hline
\multirow{2}{*}{\textbf{Model}} & \multicolumn{2}{c|}{\textbf{Tissue Seg}} & \multicolumn{2}{c|}{\textbf{Blur Seg}} & \multicolumn{2}{c|}{\textbf{Folds Seg}} & \multicolumn{2}{c|}{\textbf{Pen Seg}} \\ \cline{2-9} 
                                & Dice               & Time                & AUC               & Time               & Dice               & Time               & Dice               & Time             \\ \hline
\textbf{UNet-ENet-b0}           & 0.882              & 82.46               & 0.862             & 293.72             & 0.793              & 301.65             & 0.904              & 2.89             \\ \hline
\textbf{UNet-ResNet18}          & 0.913              & 97.06               & 0.875             & 363.56             & 0.825              & 384.66             & 0.926              & 2.93             \\ \hline
\textbf{UNet-ResNet34}          & 0.904              & 114.83              & 0.886             & 423.53             & 0.817              & 415.72             & 0.912              & 3.16             \\ \hline
\textbf{UNet++-ENet-b0}         & 0.891              & 215.73              & 0.843             & 781.63             & 0.782              & 841.72             & 0.892              & 4.71             \\ \hline
\textbf{UNet++-ResNet18}        & 0.911              & 260.63              & 0.879             & 827.73             & 0.812              & 861.84             & 0.924              & 5.06             \\ \hline
\textbf{UNet++-ResNet34}        & 0.914              & 320.68              & 0.893             & 1049.47            & 0.835              & 974.47             & 0.947              & 5.62             \\ \hline
\end{tabular}

\caption{Performance and inference times of different variants of four segmentation models in our study. Dice score on TCGA-Foreground dataset, AUC-ROC on FocusPath dataset and Dice scores on hand annotated tissue folds and pen marker segmentation dataset along with inference time (in seconds) on randomly selected 10 WSIs from TCGA dataset.}
\label{table_seg_model_compare}
\end{table}

\end{document}